# Hedging: Scaling and the Investor Horizon*

## John Cotter and Jim Hanly

John Cotter,
Director of Centre for Financial Markets,
School of Business
University College Dublin,
Blackrock,
Co. Dublin,
Ireland,
Tel 353 1 716 8900,
e-mail john.cotter@ucd.ie.

Jim Hanly,
School of Accounting and Finance,
Dublin Institute of Technology,
Dublin 2,
Ireland.
tel 353 1 402 3180,
e-mail james.hanly@dit.ie.

*The authors would like to thank participants at a University College Dublin seminar for their comments on an earlier draft. Part of this study was carried out while Cotter was visiting the UCLA Anderson School of Management and is thankful for their hospitality. Cotter's contribution to the study has been supported by a University College Dublin School of Business research grant. The authors thank an anonymous referee for helpful comments but the usual caveat applies.

# Hedging: Scaling and the Investor Horizon

## Abstract

This paper examines the volatility and covariance dynamics of cash and futures contracts that underlie the Optimal Hedge Ratio (OHR) across different hedging time horizons. We examine whether hedge ratios calculated over a short term hedging horizon can be scaled and successfully applied to longer term horizons. We also test the equivalence of scaled hedge ratios with those calculated directly from lower frequency data and compare them in terms of hedging effectiveness. Our findings show that the volatility and covariance dynamics may differ considerably depending on the hedging horizon and this gives rise to significant differences between short term and longer term hedges. Despite this, scaling provides good hedging outcomes in terms of risk reduction which are comparable to those based on direct estimation.

# Hedging: Scaling and the Investor Horizon

## I        Introduction

Much of the large body of work on hedging has focused on short time horizons such as a 1-day frequency[1]. This ignores the fact that the OHR is dependent on the time horizon and that a hedge calculated from data at one frequency may not provide good risk reduction outcomes over lower frequencies. Also, there is no definitive answer to the question of what constitutes the relevant horizon for risk management since investors have differing investment horizons. Research suggests a variety of horizons ranging from 1-day for traders, to 1-month or even 12 months for investors and corporate risk management respectively.[2] However, the estimation of hedge strategies over longer term time horizons such as weekly or monthly can be problematic. The key problem is that for lower frequency data, fewer observations are available on which to base the estimation (Hwang and Valls Pereira, 2006). For example, an estimation period of 5-years of daily data will yield around 1300 observations; however, this number drops to around 260 for the weekly frequency and just 60 if data at the monthly frequency are used. Using a longer estimation period such as 20 years may not yield better estimates since OHR's are time varying (Lien and Tse, 2002). This means that using a very long estimation period may be suboptimal, as some of the data may not be relevant given their temporal distance.[3] Also, time varying hedge strategies such as those calculated using GARCH models, require large numbers of observations if the model is to meet the non-negativity constraints which are a typical feature of those models[4]. The estimation and statistical problems related to low sample size, have led to the adoption by risk managers, of models that estimate volatility at one frequency (say daily) and then scale the volatility estimate to obtain low frequency volatilities (say monthly). A number of scaling laws have been applied in financial applications. Initially these were based on Gaussian distributions and random walks which allowed scaled sums of Gaussians to follow the same distribution. Further work by Mandelbrot (1963) proved that daily prices were not from a Gaussian distribution and a number of papers, including Dacarogna, Muller, Pictet and De Vries (1998) provided evidence of scaling and power laws in many financial time series[5]. Scaling has been used in many applications in financial economics, most notably in the Black-Scholes-Merton option pricing framework and in risk management and banking regulation, where scaling is based on the Square-Root-of-Time (SQRT) rule [6]. For example, a typical daily volatility estimate for equity index data using the standard deviation would be about 1.3%. Applying the SQRT rule to obtain a monthly estimate would yield a volatility of 5.8%, or an annual estimate of about 21%.

---

[1] See Lien and Tse (2002), for a review of the development of the literature on optimal hedging.
[2] See Locke (1999), Smithson and Minton (1996), for evidence on the variety of time horizons that are relevant to investors.
[3] See for example Merton (1980) who finds that better volatility estimates are to be had from using a large number of high frequency data rather than a small number of low frequency data over a longer time period.

While scaling has been extensively applied in the literature on volatility and high frequency finance, little work has been done on scaling and hedging. However, a number of papers have examined hedging for different time horizons (see Malliaris and Urrutia (1991), Geppert (1995), Chen, Lee and Shrestha (2004) and In and Kim (2006)), and the findings show that as the hedging horizon increases, both the OHR and the in-sample hedging effectiveness increase. There are contrasting findings out-of-sample with some papers reporting decreased effectiveness at longer time horizons (Malliaris and Urrutia, 1991). However, most papers report a positive relationship between hedge horizon and hedging effectiveness (Chen et al, 2004, Lien and Shrestha, 2007). A variety of methods have been used to calculate hedges over longer time horizons. These include the use of models based on the underlying data generating process and more recently the use of wavelet analysis[7].

---

[4] See Hwang and Valls Pereira (2006) for a discussion.
[5] For an extensive discussion on scaling laws in finance see Brock (1999).
[6] The Basel agreements on banking supervision require that daily volatility estimates be scaled using the SQRT rule.

In this paper we calculate a hedge ratio at a relatively high frequency using a 1-day (daily) time horizon. We then apply this hedge ratio by scaling it up to both 5-day (weekly) and 20-day (monthly) frequencies. We compare the scaled hedge ratios with OHR's that are estimated by matching the frequency of the data to the hedging horizon[8]. We also re-examine the issue of hedging effectiveness across different time horizons using the performance evaluation criteria of Value at Risk (VaR) and Conditional Value at Risk (CVaR). To our knowledge, these criteria have not been applied in the literature on hedging to evaluate hedges over different time horizons. Indeed the findings on the relationship between hedging effectiveness and time horizon are based on a single evaluation criterion which is variance. By applying additional performance evaluation criteria, we will provide additional evidence on the issue of hedging effectiveness across different time horizons.

A further contribution of this paper is an analysis of the levels of persistence in volatility for different hedging horizons. Drost and Nijman (1993), Diebold et al (1998), and Christofferson, Diebold and Schuermann (1998), all address the issue of how the accuracy of volatility forecasts change with the time horizon. The findings of these papers show that forecastability will decrease as we move from short to long time horizons. We investigate the implications of this finding for optimal hedging by examining the temporal aggregation properties of GARCH models that have been successfully used to estimate time varying OHR's.  This allows us to draw a link between the literature on volatility persistence and hedging performance.

---

[7] For example, Geppart (1995) use a Data Generating Process to generate returns at different time horizons while In and Kim (2006) apply wavelets to decompose the variance and covariance over different time scales.

[8] OHR's estimated in this way are hereafter referred to as actual hedges or equivalently 'direct estimation'.

Our findings show that actual hedge strategies statistically outperform scaled hedge strategies; however the differences are only marginal when viewed from an economic perspective. Furthermore, we find that scaled hedges are effective in that they provide acceptable reductions in risk as measured by Variance, VaR and CVaR. We also provide evidence that ex-post hedging effectiveness increases as we move from high to low frequency hedging.  Finally, we show that lower levels of volatility persistence, does not materially affect the ex-post hedging effectiveness at lower frequencies implying that GARCH models can still provide good forecast outcomes over longer hedging horizons.

The remainder of the paper proceeds as follows. In section II we outline the hedging models. In Section III we detail the scaling approach. In Section IV we describe the metrics for measuring hedging effectiveness. Section V describes the data followed by our empirical findings in Section VI. Finally, Section VII summarises and concludes.

## II    Hedging Models

Hedging using futures involves combining a futures contract with a cash position in order to reduce the risk of a position. The OHR is the ratio that minimises the risk of the payoff of the hedged portfolio which is given by:

$$+ r_{st} - \beta_t \, r_{ft} \qquad \qquad (1)$$

where $r_{st}$ and $r_{ft}$ are the returns on the cash and futures respectively, $\beta_t$ is the estimated OHR and $t$ is the subscript denoting time. We use two different hedging models in this study to estimate $\beta_t$. The first method is an OLS regression of the cash on the futures returns which yields the minimum variance hedge ratio (MVHR). This method has been applied extensively in the literature since (Ederington, 1979) and has been found to yield reasonably good performance. Its key advantage is its simplicity and ease of estimation. This is given as:

$$r_{st} = \alpha + \beta_t r_{ft} + \varepsilon_t \qquad \qquad (2)$$

where $\beta_t$ is the MVHR. This can also be calculated as the covariance between the cash and futures return divided by the variance of the futures return. A criticism of this approach is that the OLS HR is effectively constant whereas is has been empirically established that volatility and correlations upon which the OHR are based are both time-varying (Bollerslev, Chou and Kroner, 1992). We therefore also estimate a time varying OHR using a multivariate GARCH model.

We use the Diagonal Vech GARCH (1,1) model proposed by Bollerslev, Engle and Wooldridge (1988). This provides a useful benchmark from which to examine hedging over different time horizons given its ability to represent the dynamics of variances and covariance's (see Bauwens et al, 2006, for example) The model is specified as follows:

$$r_{st} = \mu_{st} + \varepsilon_{st} \quad r_{ft} = \mu_{ft} + \varepsilon_{ft}, \qquad \begin{bmatrix} \varepsilon_{st} \\ \varepsilon_{ft} \end{bmatrix} \Big| \Omega_{t-1} \sim N(0, H_t) \tag{3}$$

$$H_{st} = \omega_1 + \sum_{j=1}^{m} \alpha_{s,j} \varepsilon_{st-j}^2 + \sum_{k=1}^{n} \beta_{s,k} H_{st-k}^2 \tag{4}$$

$$H_{ft} = \omega_2 + \sum_{j=1}^{m} \alpha_{f,j} \varepsilon_{ft-j}^2 + \sum_{k=1}^{n} \beta_{f,k} H_{ft-k}^2 \tag{5}$$

$$H_{sft} = \omega_3 + \sum_{j=1}^{m} \alpha_{sf,j} \varepsilon_{st-j} \varepsilon_{ft-j} + \sum_{k=1}^{n} \beta_{sf,k} H_{sft-k} \tag{6}$$

where $\Omega_{t-1}$ is the information set at time $t-1$, $\varepsilon_{st}, \varepsilon_{ft}$ are the residuals, $H_{st}, H_{ft}$ denotes the variance of cash and futures and $H_{sft}$ is the covariance between them. $\omega = (\omega_1, \omega_2, \omega_3)$ is a 3x1 vector, and $\alpha_j = (\alpha_{s,j}, \alpha_{f,j}, \alpha_{sf,j})$ and $\beta_k = (\beta_{s,k}, \beta_{f,k}, \beta_{sf,k})$ are 3x1 vectors. The model contains 3 +3m+3n parameters. The matrices $\alpha_j$ and $\beta_k$ are restricted to be diagonal. This means that the conditional variance of the cash returns depends only on past values of itself and past values of the squared innovations in the cash returns. The conditional variance of the futures returns and the conditional covariance between cash and futures returns have similar structures. Because of the diagonal restriction we use only the upper triangular portion of the variance covariance matrix, the model is therefore parsimonious, with only nine parameters in the conditional variance-covariance structure of the Diagonal VECH (1,1) model to be estimated. We estimate the GARCH models using both Maximum Likelihood and Quasi Maximum Likelihood Estimators to obtain GARCH coefficients at the different frequencies[9].

---

[9] QMLE provides more consistent estimates of GARCH coefficients for the lower frequency data given the different distributional characteristics of lower frequency data which tend to approximate normality.

## III      Scaling

The determination of the OHR requires an estimate of the variance of the futures return at whatever frequency is being examined, together with an estimate of the covariance or correlation between the cash and futures return. The problem of estimating volatility over longer term time horizons has been examined in some detail in the risk management literature. Risk managers have good high frequency data but require reliable estimates of volatility at low frequencies corresponding to the respective holding periods of investors. An alternative approach is to estimate volatility using high frequency data and then scale it to obtain low frequency estimates. The most popular method of scaling volatilities is based upon the SQRT rule. The theoretical justification and background for the SQRT rule is as follows: Consider $S_t$, the log price of an asset at time $t$, where the changes in the log price are independent and identically distributed (i.i.d.). Then the price at time $t$ can be expressed as

$$S_t = S_{t-1} + \varepsilon_t \qquad \varepsilon_{t,i} \sim \left(0, \sigma^2\right) \tag{7a}$$

and the 1-day return, is

$$r_t = S_t - S_{t-1} = \varepsilon_t \tag{7b}$$

with variance $\sigma^2$. Aggregating h-day returns results in

$$r_t = S_t - S_{t-h} = \sum_{n=0}^{h-1} \varepsilon_{t-n} = \varepsilon_{t-h+1} + ... + \varepsilon_t \tag{7c}$$

with variance $h\sigma^2$ and standard deviation $\sqrt{h}\sigma$, which implies the square-root-of-time rule. This rule can be considered a special case of the more general empirical scaling law discussed by Dacarogna et al (2001), which gives a direct relation between time intervals $\Delta t$ and the average volatility as measured by a certain power $P$ of the absolute returns observed over those intervals. This is given as

$$\left\{ E[\,|r|^p\,] \right\}^{1/p} = c(p)\Delta t^{\,D(p)} \tag{8}$$

where $E$ is the expectation operator and $c(p)$ and $D(p)$ are deterministic functions of $p$. $D$ is the drift exponent which determines the scaling behaviour across different data frequencies. For $p = 2$, the standard deviation will scale according to the following rule

$$\left\{ E[\,|r|\,] \right\} = c\Delta t^{\,D} \tag{9}$$

where c is a constant depending on the underlying time series. For the Gaussian random walk model the drift exponent is $D = 0.5$ which yields the SQRT rule for scaling volatility. This scaling law has been widely applied both by practitioners and academics to obtain scaled volatility estimates for use in option pricing via the Black-Scholes-Merton model (eg. see Hull, 2008, chapter 13) where the h-period volatility is given by $\sigma\sqrt{h}$. It has also been widely used for estimating quantiles and in particular for risk measures such as VaR. (see Danielsson and Zigrand, 2006). For example, the 1-day VaR can be multiplied by $\sqrt{10}$ to obtain the 10-day VaR. This method of scaling by the SQRT rule has been widely used within the financial services industry as recommended by the Basel Committee on Banking Supervision (2004). It is broadly used because it is easy to understand and apply and because there are no simple alternatives. However, there are a number of objections to the use of the SQRT rule. In the first instance it requires the assumption that the log returns are i.i.d. However, high frequency financial returns are not i.i.d as evidenced by the numerous papers documenting strong volatility persistence in financial returns[10]. Also, because the SQRT rule magnifies volatility fluctuations when they should be damped, it tends to produce overestimates of long-horizon volatility (see Diebold et al, 1998 for a discussion).

---

[10] See for example the survey on ARCH and GARCH effects in Bollerslev, Chou and Kroner (1992).

Despite these issues, the SQRT method has been widely adopted in risk management and banking as a means of obtaining low frequency volatilities. Dowd and Oliver (2006) provide limited support for the use of the SQRT rule. They suggest that the key to using the SQRT rule is to apply it to the unconditional volatility as opposed to the most recent estimate[11]. Furthermore, Anderson et al (2001) report the means of the popular measure realized variance estimators grow linearly with time, which would be consistent with the SQRT rule. Dacarogna et al (2001) find that this scaling law is appropriate for a wide range of financial data and for time intervals ranging from 10 minute to more than a year. They also estimate scaling exponents for foreign exchange data and find values of $D$ very close to 0.5 for USD/GBP and other exchange rates. Additional support for the SQRT rule is put forward by Brummelhuis and Kaufman (2007) who apply it for scaling quantiles and conclude that it provides reasonable estimates for risk management purposes. Therefore, the use of the SQRT rule should provide a good approximation in terms of converting volatility from one frequency to another frequency even where the distribution is not strictly normal.

An alternative to scaling volatility by time is the use of formal model based aggregation as proposed in Drost and Nijman (1993) (henceforth DN), who study the temporal aggregation of GARCH processes. They propose the following approach: Suppose we

---

[11] This finding is based on a simulations based approach which indicate that if the daily volatility used as the basis of extrapolation is greater than the unconditional volatility it results in the SQRT rule overestimating the GARCH or 'true' volatility and vice versa. Therefore they conclude that the SQRT rule is appropriate only where the daily volatility to be used as the basis for extrapolation is equal to unconditional volatility.

begin with have a sample path of one-day returns $\left\{R_{(1)t}\right\}_{t=1}^{T}$ which follow a weak GARCH (1, 1) process as follows:

$$
\begin{aligned}
R_t &= \sigma_t \varepsilon_t \\
\sigma_t^2 &= \omega + \alpha y_{t-1}^2 + \beta \sigma_{t-1}^2 \\
\varepsilon_t &\sim NID(0,1)
\end{aligned}
\tag{10}
$$

for $t = 1, \ldots, T$ where $\sigma_t^2$ is the variance and $\omega, \alpha, \beta$ are the estimated parameters on the GARCH model. The following stationarity and regularity conditions are imposed, $0 < \omega < \infty$, and $\alpha \geq 0, \beta \geq 0, \alpha + \beta < 1$. Then DN show that, under regularity conditions, the corresponding sample path of h-day returns, $\left\{R_{(h)t}\right\}_{t=1}^{T/h}$ also follows a GARCH (1, 1) process with

$$
\sigma_{(h)t}^2 = \omega_{(h)} + \alpha_{(h)} y_{(h)t-1}^2 + \beta_{(h)} \sigma_{(h)t-1}^2
\tag{11}
$$

where

$$
\omega_{(h)} = h\omega \frac{1 - (\beta + \alpha)^h}{1 - (\beta + \alpha)}
\tag{12}
$$

$$
\alpha_{(h)} = (\beta + \alpha)^h - \beta_{(h)},
\tag{13}
$$

and $\left|\beta_{(h)}\right| < 1$ is the solution of the quadratic equation,

$$
\frac{\beta_{(h)}}{1 + \beta_{(h)}^2} = \frac{a(\beta + \alpha)^h - b}{a\left(1 + (\beta + \alpha)^{2h}\right) - 2b},
\tag{14}
$$

where

$$
\begin{aligned}
a &= h(1-\beta)^2 + 2h(h-1)\frac{(1-\beta-\alpha)^2(1-\beta^2-2\beta\alpha)}{(\kappa-1)\left(1-(\beta+\alpha)^2\right)} \\
&\quad + 4\frac{\left(h-1-h(\beta+\alpha)+(\beta+\alpha)^h\right)(\alpha - \beta\alpha(\beta+\alpha))}{1-(\beta+\alpha)^2}
\end{aligned}
\tag{15}
$$

$$
b = (\alpha - \beta\alpha(\beta+\alpha))\frac{1 - (\beta+\alpha)^{2h}}{1-(\beta+\alpha)^2},
\tag{16}
$$

and $\kappa$ is the kurtosis of $y_t$. This approach allows us to fit a GARCH model at one frequency (eg. 1-day) and the model coefficients obtained $\omega, \alpha$ and $\beta$ can then be substituted into the DN equations to obtain scaled coefficients at another frequency (E.g. 5-day). Also this formula can be used to convert 1-day covariance to h-day covariance if we substitute the covariance parameters from a multivariate GARCH model[12]. This approach is useful at generating parameter values at long time horizons from short time horizons. From the formulas for $\alpha_h$ and $\beta_h$, $\alpha_h \rightarrow 0, \beta_h \rightarrow 0$ as $h \rightarrow \infty$. This means that asymptotically the volatility becomes constant, and therefore the weak GARCH(1,1) process behaves in the limit like a random walk. This implies that conditional forecasts will have poor performance over long periods as predictability decreases. A number of papers have examined the usefulness of the DN approach. Diebold et al (1998) show that if the short horizon return model is correctly specified as a GARCH (1, 1) process, then the DN approach can be used for the correct conversion of 1-day to h-day volatility. Kaufman and Patie (2003) use the DN formula in a simulation based approach and conclude that it provides good parameter estimates based on daily data scaled to horizons of up to 1-month. A shortcoming of the approach is that it assumes an exact fit for the model whereas in general it is an approximation[13].

In this paper we provide further evidence on the applicability of the DN approach. We measure volatility persistence by fitting a GARCH (1, 1) model directly from the data at the relevant time horizon. We then use the DN formula to obtain scaled parameters from the 1-day data for both 5-day and 20-day horizons. This allows us to compare the volatility parameters estimated directly from the data with those obtained by scaling. This also allows us to examine the persistence of volatility across different time horizons and in so doing, to draw inferences about the relative forecasting ability of GARCH models across different time horizons. Our ex-ante expectation is that volatility persistence will decrease with the hedging horizon, and that this may affect the ex-post forecasting performance of hedges at lower frequencies.

---

[12] Therefore the DN approach can be used to scale hedge ratios which are composed of both variance and covariance components.
[13] See Christoffersen, Diebold, & Schuermann (1998) for a detailed discussion of the conditions under which temporal aggregation formulae may by used.

The determinants of the optimal hedge are the covariance between the cash and futures return and the variance of the futures return. Similar to the variance, the covariance can also be scaled by time under the i.i.d and normality assumptions. However, whatever the scaling method applied, the composition of the OHR effectively means that both the numerator and the denominator are scaled by the same factor. This means that when we scale the OHR, we are effectively applying a hedge ratio calculated at one frequency to hedges applied at a different frequency. We now turn to methods employed to measure the effectiveness of our hedges.

## IV    Hedging Effectiveness

We compare hedging effectiveness using the percentage reduction in the variance of the cash (unhedged) position as compared to the variance of the hedged portfolio. This is given as:

$$\% \text{ Variance Reduction} = 1 - \left[ \frac{VARIANCE_{HedgedPortfolio}}{VARIANCE_{UnhedgedPortfolio}} \right] \tag{17}$$

This measure of effectiveness has been used in the literature on hedging over different time horizons, however hedgers may seek to minimise some measure of risk other than the variance. For this reason, we use two additional hedging effectiveness metrics that will allow us to compare the hedging performance of scaled hedges with hedges calculated with data matched to the time horizon. While these metrics have been applied before to the hedging problem (see Cotter and Hanly, 2006), they have not been used to examine the relationship between time horizon and hedging effectiveness.

The second hedging effectiveness metric is VaR. For a portfolio this is the loss level over a certain period that will not be exceeded with a specified probability. The VaR at the confidence level $\alpha$ is

$$VaR_\alpha = q_\alpha \tag{18}$$

where $q_\alpha$ is the relevant quantile of the loss distribution. The performance metric employed is the percentage reduction in the VaR of the hedged as compared with the unhedged position.

$$HE_2 = 1 - \left[ \frac{VaR_{1\% HedgedPortfolio}}{VaR_{1\% UnhedgedPortfolio}} \right] \tag{19}$$

The third hedging effectiveness metric is the CVaR[14] which is the average of the worst $(1 - \alpha)100\%$ of losses.

In effect this means taking the average of quantiles in which tail quantiles are equally weighted and non-tail quantiles have zero weight.[15] The performance metric we use to evaluate hedging effectiveness is the percentage reduction in CVaR as compared with a no hedge position.

---

[14] This is also called the Expected Shortfall.
[15] For more detail on the properties of the CVaR see Cotter and Dowd (2006)

$$HE_4 = 1 - \left[ \frac{CVaR_{1\% \, HedgedPortfolio}}{CVaR_{1\% \, UnhedgedPortfolio}} \right] \qquad (20)$$

Both the VaR and the CVaR are measures of economic or monetary risk given that they provide explicit measures of the potential money loss on a portfolio as well as a probability. None of the studies that have examined hedging over different time horizons have used hedging effectiveness metrics other than the variance; therefore in applying both VaR and CVaR, we augment earlier studies and add some new findings to the literature on hedging and time-horizon.

## V    Data Description

We examine hedging over different time horizons using equity index (FTSE100), commodity (Crude Oil)[16] and foreign exchange (USD vs GBP) data. Cash and futures closing prices were obtained from Datastream. Three different frequencies were examined; 1-day (daily), 5-day (weekly) and 20-day (monthly). In each case, the returns were calculated as the differenced logarithmic prices over the respective frequencies. We obtain hedge ratios and hedged portfolios for each frequency based on direct estimation. We also obtain hedge ratios and hedged portfolios for both 5-day and 20-day by using the SQRT rule to scale the variance and covariances as described in section III. The hedged portfolios obtained in this way are labelled as scaled to distinguish them from the hedging portfolios calculated from data over the relevant interval.

---

[16] The Crude Oil contract used is the West Texas Intermediate Light Sweet Crude which trades on the Nymex in New York. This is the dominant contract for Oil trading.

Because we are examining low frequency data, we required a dataset that would be large enough to allow us to be able to fit the time varying GARCH model[17] and to carry out estimation with a reasonable degree of statistical accuracy. The full sample runs from March 29, 1993 through March 6, 2008. The estimation sample runs from March 29, 1993 through March 17, 2003 or about 10 years of data. This provides us with 2601 1-day, 521 5-day and 131 20-day observations. The remaining observations were used as a hold out sample.

[INSERT TABLE I HERE]

Descriptive statistics for the data are displayed in Table I. The following properties of the data are worth noting. As we move from the 1-day frequency to lower frequencies, the mean and volatility, as measured by the standard deviation increase. One of the more important results from a hedging perspective is that the correlation between cash and futures increases as we move from high to low frequency[18]. Also, correlations at the 1-day frequency are very different for the different assets. For example, the correlation is 0.970 for the FTSE100 but just 0.876 for Oil and 0.816 for USD/GBP. When we look at the 20-day frequency, however, there is little difference between the correlations for the different assets which are 0.992, 0.993 and 0.989 for FTSE100, Oil and USD/GBP respectively. This means hedging effectiveness should increase as we hedge longer time horizons, but also that for shorter time horizons there may be significant differences between the different assets in terms of hedging effectiveness. For this reason, the hedging model choice may be more important at the 1-day frequency.

---

[17] Hwang and Valls Pereira (2006) in an examination of the small sample properties of GARCH models report that very small numbers of observations can cause unreliable parameter estimates. Our dataset is large enough to generate reasonable estimates from the GARCH model.

[18] This finding was discussed at length in Geppert (1995) who pointed out that given a cointegrated relationship between spot and futures prices which is made up of both permanent and transitory components, over longer horizons the permanent component ties the futures and spot together while the effect of the transitory component becomes negligible.

The distribution of the data is significantly non-normal at the 1-day interval whereas at lower frequencies the data are more symmetric. For both Oil and USD/GBP, the data can be characterised as Gaussian at the 20-day frequency, as we fail to reject the hypothesis of normality at conventional significance levels. This implies that scaling using the SQRT rule may be more appropriate using lower frequency data[19]. We find significant ARCH effects at the 1% level at the 1-day frequency for each of the assets with the exception of USD/GBP which is significant at the 5% level, however, these diminish at lower frequencies. We can also observe a difference between the characteristics of the FTSE100 as compared with both the Oil and USD/GBP data. For example the FTSE100 exhibits significant ARCH effects across each hedging horizon (p-values of 0.01 or lower) whereas the Oil and USD/GBP data which have significant ARCH effects at the 1-day frequency only, have insignificant ARCH effects at the 5-day and 20-day frequencies. This finding agrees with the well known result from the literature that volatility persistence decreases as we move from high to low frequency data (see, for example, Poon and Granger, 2003). The implication is that GARCH models will generate better forecasts at shorter time horizons. However, in this paper we are estimating the ex-post hedge ratio using 1-step-ahead forecasts so it remains to be seen whether the lower volatility persistence at lower frequency will affect the efficiency of the hedges[20]. Stationarity is tested using both the Phillips Peron and

---

[19] This finding means that scaling may be more applicable from say 20-Day to 1-Year than from 1-Day to 20-Day however the problem of having enough observations at even the 20-Day interval remains.

[20] Christofferson and Diebold (2000) find that volatility forecasts are unreliable beyond 10 days. However they used daily data which means in effect that volatility is forecastable up to 10-steps. In this study although we are generating a forecasts of up to 20-days ahead for the 20-Day horizon, it is based on a 1-step forecast.

Kwiatkowski, Phillips, Schmidt and Shin (KPSS) tests[21]. The results of both tests indicate that the log returns series is stationary irrespective of the sampling frequency. This indicates that the OLS estimates should be reliable across each of the time horizons.

## VI    Empirical Findings

[INSERT TABLE II HERE]

Table II reports the GARCH parameters using direct estimation at the 1-day, 5-day and 20-day frequencies together with the scaled parameters estimated from the 1-day data using the Drost Nijman approach. Examining first the results of direct estimation, as we move from high frequency to low frequency the level of persistence in volatility is significantly reduced. This is most pronounced for the Oil and USD/GBP data. For example, as we move from 1-day to 5-day, volatility persistence[22] for the FTSE drops just over 1% from 0.986 to 0.975 for cash[23]. In the case of Oil, persistence decreases by about 16% from 0.696 to 0.528 while for USD/GBP the drop is even more pronounced going from 0.853 to 0.554. When we examine the results for the 20-day frequency, persistence is still relatively high for the FTSE at 0.875, whereas for both Oil and USD/GBP it is in the region of just 0.10. These findings are supportive of previous work such as Christofferson Diebold and Schuermann (1998) that have found that volatility persistence decreases as the time horizon increases, or alternatively as we move from high frequency to low frequency data. The implications of this relate to the forecasting ability of the GARCH models at lower frequency, where we may see a performance differential in the ex-post hedging effectiveness of the hedges for the different assets at different frequencies.

---

[21] For brevity we have only included the results from the KPSS test in Table I.
[22] As measured by the sum of the A or ARCH coefficient and the B or GARCH coefficient.
[23] For brevity we base the parameter comparisons on the cash asset unless otherwise stated.

We also report the scaled parameters obtained using the DN approach based on inputs from the GARCH model at the 1-day frequency. Comparing first the scaled parameters for the FTSE, they are broadly in line with the parameters obtained from direct estimation at the 5-day frequency but when we move to the 20-day frequency larger differences emerge. For example, the scaled parameters at the 5-day frequency are lower by about 2% – 3% as compared with the 5-day actual. When we move to the 20-day frequency, however, there are significant differences. For example, persistence as measured using the scaled coefficients at 0.761 is significantly lower than the actual estimate of 0.875 would suggest. For both Oil and USD/GBP, the scaled parameters are significantly different from the actual parameters at both the 5-day and 20-day frequencies. For Oil, actual volatility persistence at the 5-day frequency is 0.528 and 0.099 at the 20-day frequency. This compares with scaled persistence of 0.163 and just 0.0007 for the 5-day and 20-day frequencies respectively. Similarly, the actual coefficients for the USD/GBP data are very different from those obtained using the DN approach. These findings indicate that the DN approach may provide a reasonable approximation for the volatility data generating process for equity returns using a scaling factor in the region of 5, from 1-day to 5-day. This finding agrees with Kaufmann and Patie (2003) who found that the DN approach provided reasonable estimates of coefficients when scaling from 1-day frequency up to 1-week. However for other assets such as Oil or foreign exchange rates, the coefficient estimates obtained by scaling do not approximate those obtained by direct estimation.

[INSERT TABLE III HERE]

The estimated optimal hedges are presented in Figures 1a, 1b and 1c together with their associated statistics in Table III. All hedges are less than 1 with the exception of the USD/GBP at the 20-day frequency[24]. For the FTSE and USD/GBP, the OHR's increase in line with the time horizon and tend to approach the Naïve hedge ratio which is 1. For example, the mean OHR for the FTSE goes from 0.875 to 0.934 to 0.957 as we move from the 1-day frequency to the 5-day and 20-day frequencies respectively. This means that the investor would sell a larger number of futures contracts to achieve the optimal hedge as the frequency of the hedge increases. This would have implications on the cost of the hedging strategy. For example, it would mean that hedging a single 20-day period would be more expensive than a single 1-day period.

For Oil, the OHR's increase as we move from the 1-day to the 5-day horizon from 0.929 to 0.992. However, while there is then a slight decrease with the 20-day OHR at 0.987. the results are broadly similar to those for the FTSE and USD/GBP. Also the mean OHR's for the different time horizons are significantly different based on standard t-tests at the 1% level. These findings reflect the fact that the correlation between cash and futures increases as we move from high frequency to low frequency data.

The OLS hedges also tend to increase with the hedging horizon with the exception of Oil for the 20-day frequency. These results are consistent with the literature (see, for example, Chen Lee and Shrestha, 2004) in that for longer investment horizons, the optimal hedge ratio converges towards the Naïve hedge ratio. In terms of dispersion of the OHR's, there are also large differences in the standard deviations of the OHR's at the different time horizons. This is most pronounced for the Oil hedges. For example, the standard deviation of the 1-day time varying OHR is 0.127 but this drops to 0.053 for the 5-day OHR's and to just 0.008 for the 20-day hedge. This can be confirmed with a quick glance at figure 1b. This finding highlights the fact that at we move from high to low frequency hedges, the gap between the time-varying and the constant hedges narrows.

---

[24] A hedge ratio in excess of 1 implies that an investor hold a naked short position in futures which in some cases would be necessary in order to achieve the minimum variance portfolio.

We now turn to the hedging effectiveness of the hedge strategies and compare the effectiveness of hedges calculated directly using data at the relevant hedging horizon with the hedges based on scaling. Table IV presents the hedging effectiveness metrics for both OLS and GARCH models across hedging horizons. Hedging models are compared in terms of the percentage reduction in a given risk measure as compared with a no-hedge position. The effect of hedging horizon on hedging effectiveness is apparent in that in-sample hedging effectiveness increases with the hedging horizon. For example, using the variance as the performance criterion, both the OLS and GARCH hedges for USD/GBP yield around a 67% reduction in risk at the 1-day horizon. This increases to around 90% at the 5-day horizon and around 98% at the 20-day horizon. Both FTSE and Oil exhibit similar results[25].

The differences between the hedging performances at the different time horizons are also statistically significant at the 1% level i.e. the 5-day performance is statistically significantly better than the 1-day performance, similarly the 20-day is significantly better than the 5-day. Comparisons are based on t-tests of the difference in the average performance of the different hedges. These were estimated using standard errors based on Efrons (1979) bootstrap methodology. The findings are robust in that they apply to all assets. Furthermore, both the VaR and CVaR metrics confirm the findings based on the Variance performance criterion. This supports the broad findings in the literature in relation to in-sample hedging performance at different time horizons. In addition, our findings make a further contribution in that the literature has hitherto based its results on statistical performance criterion (the variance) alone, whereas by using economic criteria such as VaR and CVaR, we find further evidence of the positive relationship between hedging effectiveness and time horizon.

---

[25] Results are rounded to two decimal places, however the similarity between the OLS and GARCH hedging performance in not surprising, given that the results are based on averages of the hedging effectiveness of individual hedges. This finding further supports Cotter and Hanly (2006) who find little difference in hedging performance between OLS and GARCH hedging strategies.

We now compare the hedging effectiveness of the scaled hedges with those based on direct estimation. Table V presents t-Statistics for differences of mean hedging effectiveness for in-sample hedges obtained from 5-day actual and 5-day scaled estimation periods and similarly for the 20-day estimation period. The null hypothesis that the hedging effectiveness is the same for actual and scaled hedges can be rejected across all assets and both hedging horizons in 89% of cases indicating that there is a statistical difference in the hedging effectiveness between actual and scaled hedges. If we are to look at just the Variance and VaR, there are differences between the actual and scaled hedges in all cases. In terms of which hedging strategy performs better, differences emerge between assets and depending on hedging horizon.

[INSERT TABLE V HERE]

For the 5-day frequency, if we use just the variance as the measure of hedging effectiveness, the actual hedges perform better for all assets. If we also take into account the VaR and CVaR, the scaled hedges tend to perform relatively better for the USD/GBP contract. Looking at the 20-day frequency, the variance measure indicates that the actual hedges perform best, however the VaR and CVaR of the scaled hedges are lower for the Oil contract. These results indicate that actual hedges tend to perform significantly better than scaled hedges. However, despite the statistical difference, it would appear that the hedging performance may not be significantly different from an economic perspective. To demonstrate this, consider that the CVaR for the 5-Day actual hedge on the FTSE estimated using the GARCH model for a $1,000,000 exposure would be $15,100. This compares with an exposure of just $16,300 if the scaled hedge were used. The difference is just $1,200 whereas the relevant t-stat is 8.94. Similar differences apply to the Oil contract however they are more pronounced for the USD/GBP contract. A similar comparison for the USD/GBP contract yields a difference in the CVaR of just $500 in favour of the scaled hedge (t-stat 18.53).

At the 20-day frequency however there are larger economic differences in the effectiveness of the actual and scaled hedges. For example, again using the CVaR yields a figure of $10,100 for the actual hedge as compared with $16,500 for the scaled hedge, which is 63% larger. This reflects both the difference in hedging strategy but also the fact that the exposure is over a longer period. The differences in hedging effectiveness for the FTSE are not so large while for Oil, the scaled hedges tend to perform better when VaR and CVaR are used although not by an economically significant amount. The implication of these findings is that using a 1-day hedge scaled up to a 5-day hedging horizon may be a relatively good solution, especially for the assets examined here. When we increase the hedging horizon to 20-days, however, differences emerge depending on the asset being examined. Scaled hedges would appear to be reasonable in performance terms for both FTSE and Oil but do not provide effective hedging solutions for the USD/GBP asset.

These comparisons have been based on in-sample results however a more stringent test would be to examine scaled hedges in an out-of-sample setting. To obtain out-of-sample hedges, we use one-step ahead forecasts of the variance and covariances obtained from the GARCH model. Tables VI and VII present the out-of-sample hedging effectiveness.

[INSERT TABLES VI AND VII]

The out-of-sample results broadly support the in-sample findings. There is a significant increase in hedging effectiveness as we move from the 1-day hedges to the 5-day and 20-day hedges. There are also large differences in terms of the different assets at the different frequencies. The FTSE has the best hedging performance at the 1-day hedge horizon however the performance of both the Oil and USD/GBP contracts increases dramatically as we move the hedge horizon up to 5-day and 20-day. This finding addresses a gap the literature by providing more evidence on out-of-sample effectiveness by using effectiveness criteria such as VaR and CVaR in addition to the variance reduction measure. The implication of these results is that hedging over longer time horizons would be a preferable strategy compared with hedging shorter time spans and rolling the hedges over. The differences in hedging effectiveness between assets at higher frequencies tend to converge at lower frequencies indicating that hedging effectiveness is broadly similar for different assets at longer hedging horizons.

Looking now at comparisons between the actual and scaled hedges for the out-of-sample hedges, we find that there are significant statistical differences in 83% of cases across assets and frequencies. At the 5-Day hedging horizon, the actual hedges outperform the scaled hedges for FTSE and USD/GBP, whereas for Oil the scaled hedges generally yield lower risk. However the differences again are only economically significant for the USD/GBP contract. When we examine the 20-Day hedges, we find that the actual hedges are the best performers for all assets but again the differences are only economically significant for the USD/GBP contract. For example, using a $1,000,000 exposure to illustrate, the difference between the CVaR of the USD/GBP hedge calculated directly is $6,800 lower than the CVaR of the scaled hedge. This represents a difference of 81% in favour of the actual hedge.

In terms of hedging models, there are generally no significant differences between the OLS and the GARCH models, irrespective of whether statistical or economical evaluation is used. The best model may change for a given asset or hedge horizon. We also expected ex-ante that the out-of-sample performance of the GARCH model would be relatively better as compared with the OLS model for shorter horizons because volatility persistence is greater for high frequency data. However there is no conclusive evidence that this is the case as the performance differential is not significantly different for different hedge horizons. Looking at the actual risk measures, both the OLS and GARCH models provide broadly similar performance. For the FTSE the GARCH model appears to have the edge. It outperforms the OLS model for seven out of nine cases across the difference frequencies and the different risk measures. The exceptions are the VaR at the 5-day frequency and the Variance at the 20-day frequency. For the Oil and the USUK assets, the performance is more even, and depends on the particular risk measure and the frequency. Taking Oil, for example, the OLS yields a marginally lower CVaR of 3.19 at the 20-day frequency as compared with 3.20 for the GARCH model. The similarity in performance of the models is not surprising given that it relates to the average performance of the hedges across time, and therefore performance differences tend to be quite small. Note also that the variance of the hedged portfolios tends to increase as we increase the hedging horizon. The only exception to this is for Oil at the 5-day and 20-day frequencies and this relates to very similar performance for these particular hedges.

## VII    Summary and Conclusion

This paper compares hedge strategies across three different investor time horizons. We calculate hedges using volatility and covariance estimates based on direct estimation and compare these with hedges obtained using scaled data. Significant differences emerge between the hedge strategies, indicating that scaled hedges tend to be lower in absolute value and less volatile than those obtained from direct estimation. These differences can be traced back to the correlation properties of cash and futures which increase as the hedging horizon lengthens. Despite these differences, scaled hedges provide good outcomes in terms of absolute hedging effectiveness across all assets.

In terms of the relative performance of scaled versus actual hedges, for Equity Index and Oil hedges, particularly when scaling 1-day hedges up to 1-week, the relative performance is broadly similar. For foreign exchange hedges especially at the 20-day frequency, the results of scaling are poor when compared with the actual hedges.  We conclude therefore that only for the foreign exchange (USD/GBP) hedges can economic and statistical differences be called significant, therefore the broad finding is that scaling provides reasonably good outcomes in reducing risk.

We also examine the temporal aggregation properties of a GARCH model using the Drost Nyman approach which allows us to compare volatility persistence for high frequency and low frequency data. The results show that lower levels of volatility persistence do not materially affect the ex-post hedging effectiveness at lower frequencies. We also provide further evidence that ex-post hedging effectiveness increases as we move from high to low frequency hedging. The implications of our findings are twofold. Firstly, scaling provides good absolute, and reasonable relative hedging outcomes vis a vis direct estimation while avoiding some of the statistical problems associated with direct estimation at low frequencies. Furthermore, it is particularly useful for assets such as equity index hedges and for relatively short time scales such as 1-day to 5-day. For time scales such as 20-day or longer, the findings seem to suggest that constant or average OHR's approach the Naïve hedge ratio of 1, and therefore it may be that this approach may prove suitable for hedges based on time horizons longer than 1-month.

## Table I: Descriptive Statistics

This table presents summary statistics for the log returns of each cash and futures series. The Mean and Standard Deviation (SD) are expressed as percentages. A comparison of the scaled and the actual data shows that as we scale the 1-day standard deviations to the 5-day frequency using the SQRT rule, the average deviation of the scaled standard deviation as compared with the actual standard deviation across all assets would be just 1.14% however at the 20-day frequency the average difference rises to 7.7%. This indicates that scaling provides a good approximation up to a factor of five but that it declines rapidly thereafter. The excess skewness statistic measures asymmetry where zero would indicate a symmetric distribution. The excess kurtosis statistic measures the shape of a distribution where a value of zero would indicate a normal distribution. The Jarque and Bera (J-B) statistic measures normality. LM with 4 lags is the Lagrange Multiplier test for ARCH effects proposed by Engle 1982 and is distributed as $\chi^2$. Stationarity is tested using the KPSS test which tests the null of stationarity against the alternative of a unit root. Critical Values for the KPSS test at the 1% level are 0.739 and 0.216 for the constant and trend statistics respectively. The correlation coefficient between each set of cash and futures is also given. Associated p-Values are given in brackets.

| | 1-Day | | 5-Day | | 20-Day | |
|---|---|---|---|---|---|---|
| | Cash | Futures | Cash | Futures | Cash | Futures |
| **FTSE 100** | | | | | | |
| **Mean** | 0.009 | 0.009 | 0.052 | 0.051 | 0.23 | 0.22 |
| **SD** | 1.11 | 1.18 | 2.47 | 2.56 | 4.67 | 4.78 |
| **SD Scaled** | | | 2.48 | 2.64 | 4.96 | 5.28 |
| **Skewness** | -0.166 | -0.086 | -0.578 | -0.496 | -0.402 | -0.405 |
| | (0.00) | (0.00) | (0.00) | (0.00) | (0.06) | (0.06) |
| **Kurtosis** | 2.87 | 2.29 | 3.12 | 2.85 | 0.71 | 0.55 |
| | (0.00) | (0.00) | (0.00) | (0.00) | (0.00) | (0.00) |
| **J-B** | 916.94 | 579.29 | 241.04 | 197.78 | 18.30 | 14.78 |
| | (0.00) | (0.00) | (0.00) | (0.00) | (0.00) | (0.00) |
| **LM** | 463.31 | 410.82 | 29.56 | 25.99 | 14.89 | 14.35 |
| | (0.00) | (0.00) | (0.00) | (0.00) | (0.01) | (0.01) |
| **KPSS - Constant** | 0.513 | 0.484 | 0.585 | 0.570 | 0.606 | 0.302 |
| **- Trend** | 0.096 | 0.092 | 0.114 | 0.113 | 0.138 | 0.139 |
| **Correlation** | 0.970 | | 0.988 | | 0.992 | |
| **OIL** | | | | | | |
| **Mean** | 0.016 | 0.016 | 0.069 | 0.066 | 0.26 | 0.26 |
| **SD** | 2.34 | 2.21 | 5.16 | 5.08 | 9.41 | 9.52 |
| **SD Scaled** | | | 5.23 | 4.94 | 10.46 | 9.88 |
| **Skewness** | -0.321 | -0.251 | -0.537 | -0.577 | -0.044 | -0.149 |
| | (0.00) | (0.00) | (0.00) | (0.00) | (0.83) | (0.48) |
| **Kurtosis** | 5.07 | 4.54 | 3.08 | 3.09 | 0.353 | 0.433 |
| | (0.00) | (0.00) | (0.00) | (0.00) | (0.42) | (0.32) |
| **J-B** | 2857.03 | 2283.23 | 231.29 | 237.14 | 0.72 | 1.51 |
| | (0.00) | (0.00) | (0.00) | (0.00) | (0.69) | (0.46) |
| **LM** | 65.03 | 43.47 | 4.39 | 3.79 | 5.74 | 6.83 |
| | (0.00) | (0.00) | (0.35) | (0.43) | (0.21) | (0.14) |
| **KPSS - Constant** | 0.063 | 0.069 | 0.066 | 0.064 | 0.071 | 0.072 |
| **- Trend** | 0.032 | 0.035 | 0.041 | 0.041 | 0.048 | 0.048 |
| **Correlation** | 0.876 | | 0.971 | | 0.993 | |
| **USD/GBP** | | | | | | |
| **Mean** | 0.003 | 0.003 | 0.011 | 0.011 | 0.040 | 0.042 |
| **SD** | 0.48 | 0.51 | 1.06 | 1.10 | 2.05 | 1.99 |
| **SD Scaled** | | | 1.07 | 1.14 | 2.15 | 2.28 |
| **Skewness** | 0.0175 | 0.0152 | -0.103 | -0.181 | -0.024 | 0.020 |
| | (0.71) | (0.74) | (0.33) | (0.09) | (0.90) | (0.92) |
| **Kurtosis** | 1.92 | 2.72 | 0.30 | 0.25 | 0.65 | 0.88 |
| | (0.00) | (0.00) | (0.15) | (0.23) | (0.13) | (0.04) |
| **J-B** | 402.53 | 812.74 | 2.95 | 4.28 | 2.32 | 4.24 |
| | (0.00) | (0.00) | (0.22) | (0.12) | (0.31) | (0.12) |
| **LM** | 18.10 | 9.18 | 14.74 | 9.67 | 3.14 | 3.44 |
| | (0.00) | (0.05) | (0.10) | (0.10) | (0.53) | (0.48) |
| **KPSS - Constant** | 0.056 | 0.056 | 0.069 | 0.072 | 0.104 | 0.111 |
| **- Trend** | 0.043 | 0.041 | 0.057 | 0.055 | 0.085 | 0.086 |
| **Correlation** | 0.816 | | 0.949 | | 0.989 | |

## Table II: GARCH (1, 1) Estimates

This table reports the maximum likelihood estimates for FTSE, Oil and USD/GBP returns for 1-Day and 5-Day frequencies for the period 29/03/1993 to 17/03/2003. Also presented are the GARCH coefficients that are based on the 1-Day parameters scaled up using the Drost Nijman formula. Volatility persistence is measured by the sum of the GARCH parameters α and β. The numbers in parentheses are robust standard errors.

$$H_{s_t} = \omega + \sum_{j=1}^{m} \alpha_j \varepsilon_{s_{t-j}}^2 + \sum_{k=1}^{n} \beta_k H_{s_{t-k}}^2,$$

$$H_{f_t} = \omega + \sum_{j=1}^{m} \alpha_j \varepsilon_{f_{t-j}}^2 + \sum_{k=1}^{n} \beta_k H_{f_{t-k}}^2$$

$$H_{sf_t} = \omega + \sum_{j=1}^{m} \alpha_j \varepsilon_{s_{t-j}} \varepsilon_{f_{t-j}} + \sum_{k=1}^{n} \beta_k H_{sf_{t-k}}$$

|  | 1-DAY | 5-DAY ACTUAL | 20-DAY ACTUAL | 5-DAY SCALED | 20-DAY SCALED |
|---|---|---|---|---|---|
| **FTSE** | | | | | |
| $\omega_s$ | 0.0000 | 0.0000 | 0.0000 | | |
| | (0.000) | (0.000) | (0.000) | | |
| $\omega_{sf}$ | 0.0000 | 0.0000 | 0.0000 | | |
| | (0.000) | (0.000) | (0.000) | | |
| $\omega_f$ | 0.0000 | 0.0000 | 0.0000 | | |
| | (0.000) | (0.000) | (0.000) | | |
| $\alpha_s$ | 0.0565 | 0.0913 | 0.2467 | 0.0746 | 0.0785 |
| | (0.004) | (0.013) | (0.081) | | |
| $\alpha_{sf}$ | 0.0561 | 0.0907 | 0.2380 | 0.0734 | 0.0747 |
| | (0.004) | (0.013) | (0.080) | | |
| $\alpha_f$ | 0.0570 | 0.0910 | 0.2318 | 0.0737 | 0.0734 |
| | (0.004) | (0.013) | (0.078) | | |
| $\beta_s$ | 0.9299 | 0.8841 | 0.6285 | 0.8594 | 0.6825 |
| | (0.004) | (0.015) | (0.090) | | |
| $\beta_{sf}$ | 0.9290 | 0.8840 | 0.6414 | 0.8546 | 0.6669 |
| | (0.004) | (0.015) | (0.091) | | |
| $\beta_f$ | 0.9272 | 0.8832 | 0.6525 | 0.8497 | 0.6535 |
| | (0.005) | (0.015) | (0.093) | | |
| **OIL** | | | | | |
| $\omega_s$ | 0.0002 | 0.0015 | 0.0000 | | |
| | (0.000) | (0.000) | (0.000) | | |
| $\omega_{sf}$ | 0.0001 | 0.0014 | 0.0000 | | |
| | (0.000) | (0.000) | (0.000) | | |
| $\omega_f$ | 0.0001 | 0.0012 | 0.0000 | | |
| | (0.000) | (0.000) | (0.000) | | |
| $\alpha_s$ | 0.2836 | 0.2569 | 0.0500 | 0.0654 | 0.0075 |
| | (0.011) | (0.042) | (0.000) | | |
| $\alpha_{sf}$ | 0.2820 | 0.2494 | 0.0500 | 0.0720 | 0.0088 |
| | (0.010) | (0.038) | (0.000) | | |
| $\alpha_f$ | 0.2780 | 0.2465 | 0.0500 | 0.1056 | 0.0204 |
| | (0.011) | (0.036) | (0.000) | | |
| $\beta_s$ | 0.4129 | 0.2717 | 0.0495 | 0.0985 | -0.0067 |
| | (0.010) | (0.028) | (0.000) | | |
| $\beta_{sf}$ | 0.4403 | 0.2853 | 0.0503 | 0.1246 | -0.0074 |
| | (0.006) | (0.024) | (0.000) | | |
| $\beta_f$ | 0.5441 | 0.3615 | 0.0494 | 0.2700 | -0.0005 |
| | (0.006) | (0.026) | (0.000) | | |

|  | 1-DAY | 5-DAY ACTUAL | 20-DAY ACTUAL | 5-DAY SCALED | 20-DAY SCALED |
|---|---|---|---|---|---|
| **USD/GBP** | | | | | |
| $\omega_s$ | 0.0000 | 0.0000 | 0.0021 | | |
|  | (0.000) | (0.000) | (0.000) | | |
| $\omega_{sf}$ | 0.0000 | 0.0000 | 0.0022 | | |
|  | (0.000) | (0.000) | (0.000) | | |
| $\omega_f$ | 0.0000 | 0.0000 | 0.0022 | | |
|  | (0.000) | (0.000) | (0.000) | | |
| $\alpha_s$ | 0.0599 | 0.1422 | 0.0545 | 0.0347 | 0.0071 |
|  | (0.007) | (0.030) | (0.000) | | |
| $\alpha_{sf}$ | 0.0824 | 0.1674 | 0.0499 | 0.0307 | 0.0037 |
|  | (0.007) | (0.032) | (0.000) | | |
| $\alpha_f$ | 0.1133 | 0.1942 | 0.0465 | 0.0328 | 0.0033 |
|  | (0.009) | (0.036) | (0.000) | | |
| $\beta_s$ | 0.7934 | 0.4119 | 0.0486 | 0.4175 | 0.0347 |
|  | (0.019) | (0.096) | (0.000) | | |
| $\beta_{sf}$ | 0.6721 | 0.3327 | 0.0497 | 0.2137 | -0.0001 |
|  | (0.021) | (0.090) | (0.000) | | |
| $\beta_f$ | 0.5858 | 0.2762 | 0.0520 | 0.1342 | -0.0025 |
|  | (0.022) | (0.086) | (0.000) | | |
| **VOLATILITY PERSISTENCE** | | | | | |
| **FTSE** | | | | | |
| $\alpha_s + \beta_s$ | 0.9864 | 0.9754 | 0.8751 | 0.9340 | 0.7611 |
| $\alpha_f + \beta_f$ | 0.9842 | 0.9742 | 0.8843 | 0.9234 | 0.7269 |
| **OIL** | | | | | |
| $\alpha_s + \beta_s$ | 0.6965 | 0.5286 | 0.0995 | 0.1639 | 0.0007 |
| $\alpha_f + \beta_f$ | 0.8221 | 0.6080 | 0.0994 | 0.3755 | 0.0199 |
| **USD/GBP** | | | | | |
| $\alpha_s + \beta_s$ | 0.8533 | 0.5541 | 0.1031 | 0.4522 | 0.0418 |
| $\alpha_f + \beta_f$ | 0.6991 | 0.4704 | 0.0985 | 0.1669 | 0.0008 |

**Table III: Optimal Hedge Ratios – Descriptive Statistics**

This table presents descriptive statistics for the time-varying GARCH Optimal Hedge Ratios at 1-Day, 5-Day and 20-Day frequencies together with statistics for the scaled hedge ratios at 5-Day and 20-Day frequencies. Stationarity is tested using the Phillips Peron unit root test with associated p-values in brackets. For all assets the OHR's are stationary. OLS Hedge Ratios are also presented.

| | ACTUAL | | | SCALED | |
| --- | --- | --- | --- | --- | --- |
| **GARCH Hedges** | **1-Day** | **5-Day** | **20-Day** | **5-Day** | **20-Day** |
| | | | | | |
| **FTSE** | | | | | |
| **Mean** | 0.875 | 0.934 | 0.957 | 0.875 | 0.881 |
| **SD** | 0.071 | 0.044 | 0.030 | 0.072 | 0.074 |
| **Minimum** | 0.687 | 0.833 | 0.885 | 0.695 | 0.695 |
| **Maximum** | 1.113 | 1.124 | 1.049 | 1.093 | 1.092 |
| **Stationarity** | -4.82 | -3.26 | -4.84 | -4.44 | -3.85 |
| | (0.00) | (0.01) | (0.00) | (0.00) | (0.00) |
| **OIL** | | | | | |
| **Mean** | 0.929 | 0.992 | 0.987 | 0.932 | 0.932 |
| **SD** | 0.127 | 0.053 | 0.008 | 0.128 | 0.132 |
| **Minimum** | -0.015 | 0.544 | 0.949 | 0.156 | 0.156 |
| **Maximum** | 1.905 | 1.167 | 1.036 | 1.710 | 1.207 |
| **Stationarity** | -23.96 | -12.38 | -13.52 | -20.36 | -10.75 |
| | (0.00) | (0.00) | (0.00) | (0.00) | (0.00) |
| **USD/GBP** | | | | | |
| **Mean** | 0.776 | 0.917 | 1.018 | 0.778 | 0.790 |
| **SD** | 0.078 | 0.049 | 0.021 | 0.078 | 0.063 |
| **Minimum** | 0.099 | 0.648 | 0.938 | 0.368 | 0.382 |
| **Maximum** | 1.006 | 1.102 | 1.092 | 1.006 | 1.006 |
| **Stationarity** | -19.62 | -14.92 | -17.77 | -18.95 | -10.20 |
| | (0.00) | (0.00) | (0.00) | (0.00) | (0.00) |
| | | | | | |
| **OLS Hedges** | | | | | |
| | | | | | |
| **FTSE** | 0.907 | 0.955 | 0.968 | 0.907 | 0.907 |
| **OIL** | 0.928 | 0.984 | 0.981 | 0.928 | 0.928 |
| **USD/GBP** | 0.756 | 0.918 | 1.016 | 0.756 | 0.756 |

**Table IV: Hedging Performance of Actual and Scaled Hedge Strategies**

This table presents in-sample hedging performance measures for hedge strategies at the 1-Day, 5-Day and 20-Day time horizons calculated from Actual data together with performance measures for hedges based on Scaled data for both the 5-Day and 20-Day horizons. The performance measures are the Variance, VaR at the 1% level and CVaR at the 1% level. Hedging effectiveness is measured as the percentage reduction in the relevant performance measure as compared with a no-hedge strategy and is reported as the figure in brackets. For example, examining the results for the FTSE using the Actual data at the 5-day horizon, the GARCH model yields a 97% (0.97) reduction in the variance of a hedging portfolio as compared with a no-hedge strategy

| | | | | IN-SAMPLE | | | | | | |
| | | | ACTUAL | | | | | | SCALED | |
| | 1-DAY | | 5-DAY | | 20-DAY | | 5-DAY | | 20-DAY | |
| | OLS | GARCH | OLS | GARCH | OLS | GARCH | OLS | GARCH | OLS | GARCH |
|---|---|---|---|---|---|---|---|---|---|---|
| **FTSE** | | | | | | | | | | |
| VARIANCE (x10$^{-4}$) | 0.074 | 0.067 | 0.145 | 0.139 | 0.363 | 0.375 | 0.160 | 0.167 | 0.446 | 0.555 |
| | (0.94) | (0.95) | (0.97) | (0.97) | (0.98) | (0.98) | (0.97) | (0.97) | (0.98) | (0.98) |
| VaR (x10$^{-2}$) | 0.83 | 0.75 | 1.08 | 1.09 | 1.57 | 1.66 | 1.16 | 1.25 | 1.76 | 1.97 |
| | (0.73) | (0.76) | (0.85) | (0.85) | (0.88) | (0.88) | (0.84) | (0.83) | (0.87) | (0.85) |
| CVaR (x10$^{-2}$) | 1.13 | 1.07 | 1.54 | 1.51 | 1.85 | 1.88 | 1.52 | 1.63 | 1.87 | 2.07 |
| | (0.72) | (0.74) | (0.83) | (0.84) | (0.87) | (0.87) | (0.84) | (0.82) | (0.87) | (0.85) |
| **OIL** | | | | | | | | | | |
| VARIANCE (x10$^{-4}$) | 1.284 | 1.319 | 1.526 | 1.539 | 1.172 | 1.174 | 1.607 | 1.776 | 1.424 | 2.400 |
| | (0.77) | (0.76) | (0.94) | (0.94) | (0.99) | (0.99) | (0.94) | (0.93) | (0.98) | (0.97) |
| VaR (x10$^{-2}$) | 3.89 | 3.95 | 4.17 | 4.07 | 2.83 | 3.04 | 4.51 | 4.55 | 2.35 | 2.34 |
| | (0.41) | (0.40) | (0.67) | (0.68) | (0.88) | (0.87) | (0.65) | (0.64) | (0.90) | (0.90) |
| CVaR (x10$^{-2}$) | 6.34 | 6.48 | 5.66 | 5.42 | 4.16 | 4.37 | 5.68 | 5.64 | 3.81 | 2.52 |
| | (0.31) | (0.30) | (0.71) | (0.72) | (0.84) | (0.83) | (0.70) | (0.71) | (0.85) | (0.90) |
| **USD/GBP** | | | | | | | | | | |
| VARIANCE (x10$^{-4}$) | 0.076 | 0.075 | 0.112 | 0.109 | 0.096 | 0.092 | 0.143 | 0.135 | 0.365 | 0.339 |
| | (0.67) | (0.67) | (0.90) | (0.90) | (0.98) | (0.98) | (0.87) | (0.88) | (0.91) | (0.92) |
| VaR (x10$^{-2}$) | 0.76 | 0.78 | 1.16 | 1.08 | 0.76 | 0.76 | 1.02 | 1.04 | 1.56 | 1.33 |
| | (0.38) | (0.36) | (0.53) | (0.57) | (0.84) | (0.84) | (0.59) | (0.58) | (0.67) | (0.72) |
| CVaR (x10$^{-2}$) | 1.07 | 1.06 | 1.28 | 1.25 | 1.01 | 1.01 | 1.21 | 1.20 | 1.63 | 1.65 |
| | (0.31) | (0.31) | (0.56) | (0.57) | (0.82) | (0.82) | (0.59) | (0.59) | (0.71) | (0.71) |

**Table V: Comparison of Actual vs Scaled Hedging Performance**

This table presents t-Statistics for difference of mean hedging effectiveness for in-sample hedges obtained from 5-Day actual and 5-Day scaled estimation periods and similarly for the 20-Day estimation period. * denotes not significant at 5% level.

$$H_O : HE_{scaled} = HE_{actual}, H_A : HE_{scaled} \neq HE_{actual}$$

| | | IN-SAMPLE | | | |
| | 5-DAY | | | 20-DAY | |
| | OLS | GARCH | | OLS | GARCH |
|---|---|---|---|---|---|
| **FTSE** | | | | | |
| **Variance** | 10.60 | 19.57 | | 6.93 | 13.37 |
| **VAR** | 5.16 | 10.05 | | 3.75 | 7.34 |
| **CVAR** | 1.57* | 8.94 | | 0.88* | 9.07 |
| | | | | | |
| **OIL** | | | | | |
| **Variance** | 4.37 | 12.16 | | 4.34 | 11.34 |
| **VAR** | 5.27 | 6.38 | | 3.04 | 6.17 |
| **CVAR** | 0.30* | 0.35* | | 2.74 | 25.07 |
| | | | | | |
| **USD/GBP** | | | | | |
| **Variance** | 120.91 | 85.44 | | 114.95 | 94.77 |
| **VAR** | 25.54 | 21.61 | | 44.00 | 37.10 |
| **CVAR** | 14.39 | 18.53 | | 65.62 | 75.67 |

**Table VI: Hedging Performance of Actual and Scaled Hedge Strategies**

This table presents out-of-sample hedging performance measures for hedge strategies at the 1-Day, 5-Day and 20-Day time horizons calculated from Actual data together with performance measures for hedges based on Scaled data for both the 5-Day and 20-Day horizons. The performance measures are the Variance, VaR at the 1% level and CVaR at the 1% level. Hedging effectiveness is measured as the percentage reduction in the relevant performance measure as compared with a no-hedge strategy and is reported as the figure in brackets. For example, examining the results for the FTSE, using the Scaled data at the 20-day horizon, the GARCH model yields a 98% (0.98) reduction in the variance of a hedging portfolio as compared with a no-hedge strategy

| | OUT-OF-SAMPLE | | | | | | | | | |
| | ACTUAL | | | | | | SCALED | | | |
| | 1-DAY | | 5-DAY | | 20-DAY | | 5-DAY | | 20-DAY | |
| | OLS | GARCH | OLS | GARCH | OLS | GARCH | OLS | GARCH | OLS | GARCH |
|---|---|---|---|---|---|---|---|---|---|---|
| **FTSE** | | | | | | | | | | |
| **VARIANCE ($\times 10^{-4}$)** | 0.031 | 0.026 | 0.052 | 0.046 | 0.137 | 0.146 | 0.080 | 0.052 | 0.174 | 0.155 |
| | (0.96) | (0.97) | (0.99) | (0.99) | (0.98) | (0.98) | (0.98) | (0.99) | (0.98) | (0.98) |
| **VaR ($\times 10^{-2}$)** | 0.51 | 0.48 | 0.64 | 0.65 | 0.95 | 0.92 | 0.75 | 0.64 | 0.85 | 0.94 |
| | (0.81) | (0.81) | (0.89) | (0.89) | (0.86) | (0.87) | (0.87) | (0.89) | (0.88) | (0.86) |
| **CVaR ($\times 10^{-2}$)** | 0.67 | 0.65 | 0.83 | 0.78 | 1.02 | 0.99 | 1.06 | 0.78 | 0.85 | 1.05 |
| | (0.79) | (0.80) | (0.89) | (0.90) | (0.88) | (0.88) | (0.86) | (0.90) | (0.90) | (0.88) |
| **OIL** | | | | | | | | | | |
| **VARIANCE ($\times 10^{-4}$)** | 0.720 | 0.728 | 1.184 | 1.239 | 1.156 | 1.141 | 1.231 | 1.207 | 1.584 | 1.305 |
| | (0.83) | (0.83) | (0.94) | (0.94) | (0.98) | (0.98) | (0.94) | (0.94) | (0.97) | (0.98) |
| **VaR ($\times 10^{-2}$)** | 2.91 | 2.98 | 3.98 | 3.94 | 3.09 | 2.91 | 3.74 | 3.89 | 3.50 | 3.30 |
| | (0.44) | (0.42) | (0.62) | (0.63) | (0.79) | (0.80) | (0.65) | (0.63) | (0.77) | (0.78) |
| **CVaR ($\times 10^{-2}$)** | 4.71 | 4.78 | 4.57 | 4.62 | 3.19 | 3.20 | 4.44 | 4.48 | 3.64 | 3.36 |
| | (0.27) | (0.26) | (0.69) | (0.69) | (0.80) | (0.80) | (0.70) | (0.70) | (0.77) | (0.79) |
| **USD/GBP** | | | | | | | | | | |
| **VARIANCE ($\times 10^{-4}$)** | 0.060 | 0.056 | 0.117 | 0.113 | 0.127 | 0.131 | 0.179 | 0.142 | 0.554 | 0.385 |
| | (0.78) | (0.79) | (0.91) | (0.92) | (0.98) | (0.98) | (0.87) | (0.90) | (0.90) | (0.93) |
| **VaR ($\times 10^{-2}$)** | 0.60 | 0.59 | 0.89 | 0.89 | 0.68 | 0.76 | 1.13 | 1.05 | 1.37 | 1.15 |
| | (0.56) | (0.57) | (0.65) | (0.65) | (0.85) | (0.84) | (0.55) | (0.58) | (0.71) | (0.75) |
| **CVaR ($\times 10^{-2}$)** | 0.91 | 0.89 | 1.12 | 1.11 | 0.84 | 0.83 | 1.37 | 1.27 | 1.52 | 1.32 |
| | (0.42) | (0.44) | (0.67) | (0.67) | (0.83) | (0.83) | (0.59) | (0.62) | (0.69) | (0.73) |

**Table VII: Comparison of Actual vs Scaled Hedging Performance**

This table presents T-Statistics for difference of mean hedging effectiveness for out-of-sample hedges obtained from 5-Day actual and 5-Day scaled estimation periods and similarly for the 20-Day estimation period. * denotes not significant at 5% level.

$$H_O : HE_{scaled} = HE_{actual}, H_A : HE_{scaled} \neq HE_{actual}$$

| | OUT-OF-SAMPLE | | | |
| | 5-DAY | | 20-DAY | |
| | OLS | GARCH | OLS | GARCH |
|---|---|---|---|---|
| **FTSE** | | | | |
| **Variance** | 28.47 | 9.67 | 6.55 | 1.41* |
| **VAR** | 9.29 | 0.20* | 4.09 | 0.88* |
| **CVAR** | 14.38 | 0.23* | 9.71 | 3.93 |
| | | | | |
| **OIL** | | | | |
| **Variance** | 2.57 | 1.76* | 7.09 | 3.16 |
| **VAR** | 4.86 | 1.10* | 3.93 | 4.10 |
| **CVAR** | 2.86 | 2.99 | 11.66 | 4.04 |
| | | | | |
| **USD/GBP** | | | | |
| **Variance** | 184.93 | 121.44 | 102.07 | 57.17 |
| **VAR** | 45.94 | 39.47 | 40.07 | 17.94 |
| **CVAR** | 36.64 | 41.67 | 68.48 | 56.23 |

Figure 1a: Optimal Hedge Ratios

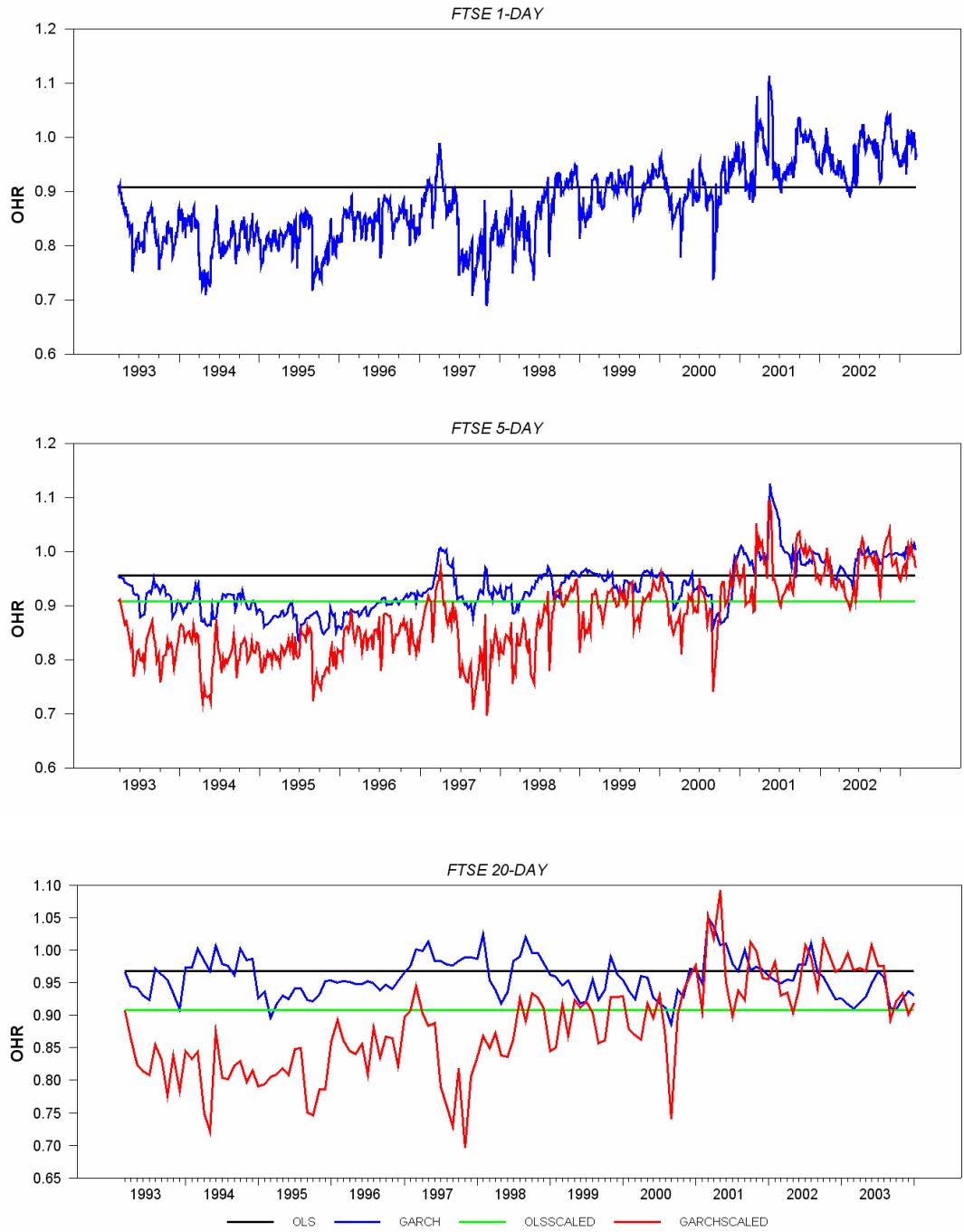

Figure 1b:    Optimal Hedge Ratios

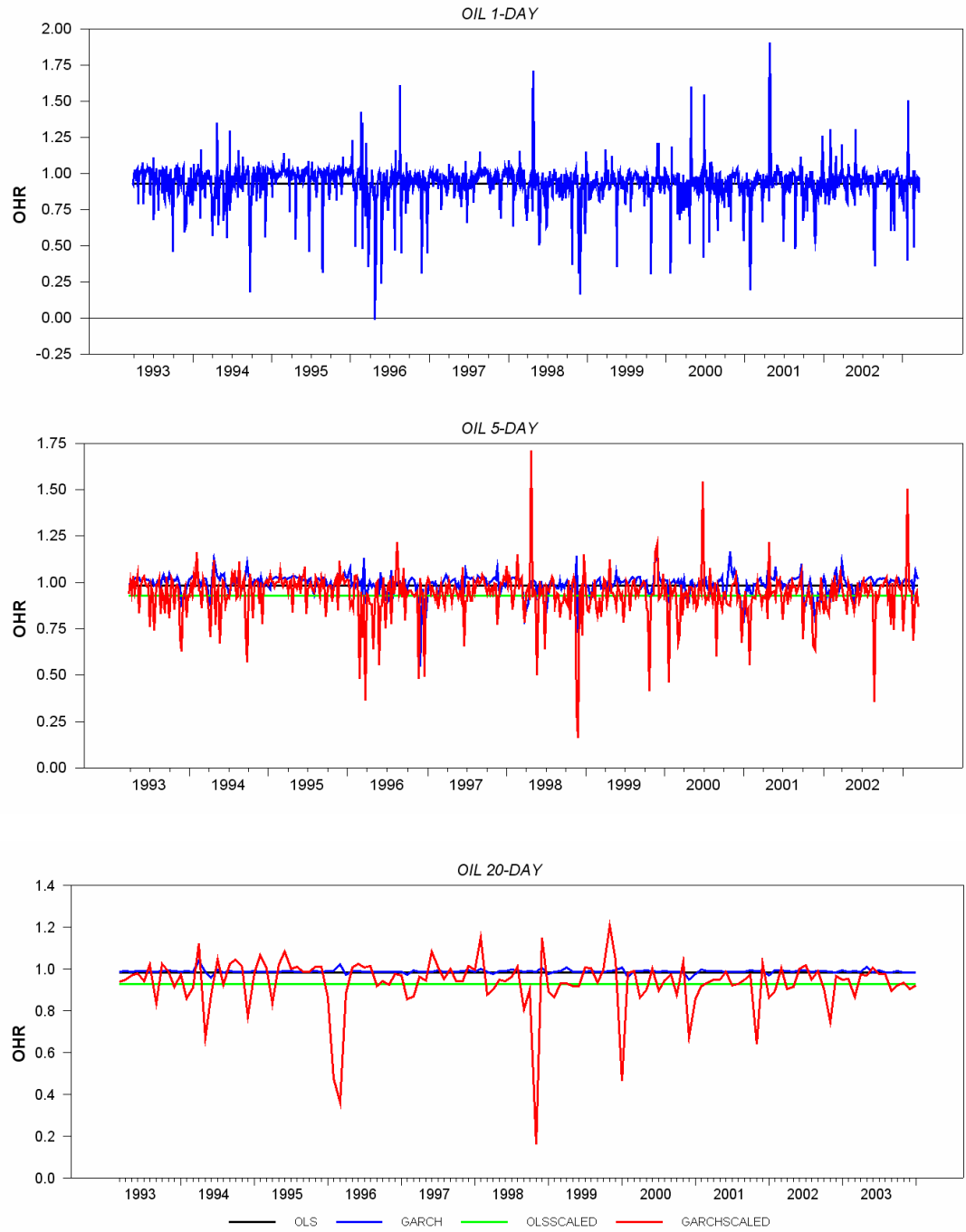

Figure 1c:        Optimal Hedge Ratios

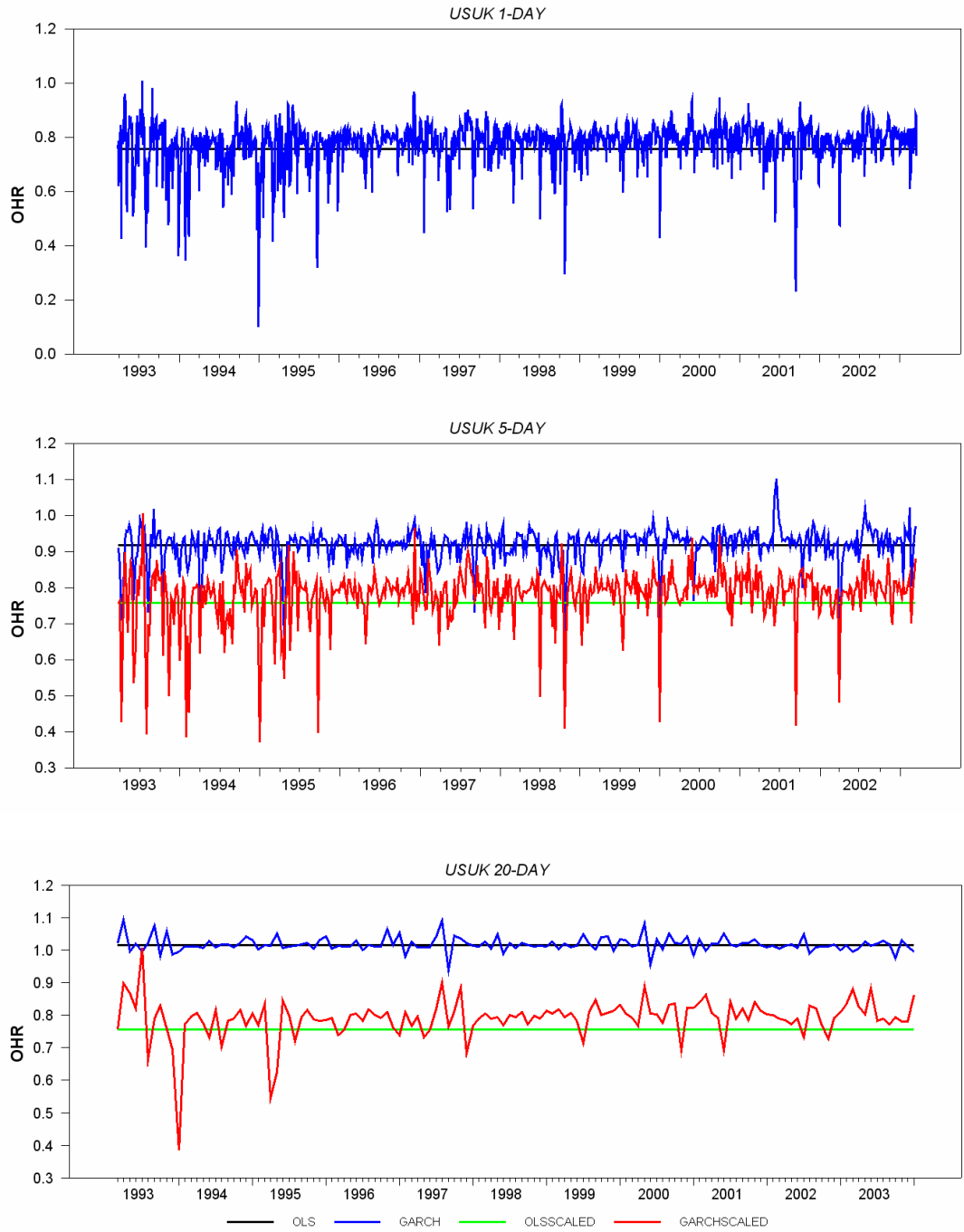